\documentstyle[referee]{laa}

\begin{document}

\thesaurus{10(01.03.1; 01.05.1; 03.02.1; 07.37.1; 18.10.1)}

\title {The Nordtvedt effect in the Trojan asteroids}

\author {R.B. ORELLANA\inst{1,*}\and H. VUCETICH\inst{2,*}}

\institute {Observatorio Astron\'omico de La Plata, Paseo del Bosque,
 (1900)La Plata, Argentina
\and Departamento de F\'isica, Universidad Nacional de La PLata, C.C. 67,
 (1900)La Plata, Argentina}
\thanks{Member of C.O.N.I.C.E.T. (Argentina)}

\date{Received October 5, accepted December 15, 1992}

\maketitle
\begin{abstract}
Bounds to the Nordtvedt parameter are obtained from the motion of the first
twelve Trojan asteroids in the period 1906-1990. From the analysis performed,
we derive a value for the inverse of the Saturn mass 3497.80$\pm$0.81 and the
Nordtvedt parameter -0.56$\pm$0.48, from a simultaneous solution for all
asteroids.
\keywords{relativity - gravitation - asteroids - astronomical constants -
celestial mechanics}
\end{abstract}

\section{Introduction}
           The asteroids located in the vicinity of the equilateral triangle
solutions of Lagrange (L$_4$ and L$_5$), known as the Trojan asteroids, are
particularly sensitive to a possible violation of the Principle of
Equivalence (Nordtvedt, 1968) because they act as a resonator selecting long
period perturbations. Theories of gravitation alternatives to General
Relativity predict a difference between inertial $({\em{m_i}})$ and passive
gravitational $({\em{m_g}})$ masses of a planetary-sized body (the so-called
Nordtvedt effect) equal to:
\begin{equation}
{{m_g}\over{m_i}}  =  1 + \Delta,           \label{Omega}
\end{equation}

            For the sun, the correction term $\Delta$ is equal to:
\begin{equation}
\Delta_\odot = {{15 G M_\odot}\over{{2 R_\odot}c^2}}\eta,
\end{equation} 
$M_\odot$ is the solar mass, $R_\odot$ is the solar radius and $\eta$, the
Nordtvedt parameter, is a linear function of the PPN parameters (Will,
1981). It can be shown that the null result that General Relativity predicts
comes from the cancellation of several large contributions to $\Delta$; and
thus a null result for the Nordtvedt effect becomes a strong test of General
Relativity.

             The standard value of $\Delta_\odot$ is
\begin{equation}
\Delta_\odot = 1.59\;10^{-5}\;\eta.
\end{equation}

           For $\eta\simeq1$, this term produces a shift of the Lagrange
points, L$_4$ and L$_5$, toward Jupiter by an amount of approximately one
arcsecond.  This effect is very large and easily discerned by standard
astrometric techniques if systematic errors can be controlled in the
photographic observations.

           In 1988, we realized a first determination of the Nordtvedt
parameter using the first six Trojan asteroids (Orellana-Vucetich, 1988), and
we obtained a value of $\eta = 0.0\pm$0.5, in agreement with the General
Relativity prediction. In this paper, we incorporate six new asteroids and
we increase the span of observations. The enlarged set permits a rigorous
analysis of systematic errors, specially biases in the star catalogue
reference system.

           In the following sections, we shall discuss the new results.

\begin{table}
\caption{Observational set parameters. The columns show the number and name
of the asteroid, time span covered by observations, accepted number and the
number of the libration point.}
\label{Param}
\begin{flushleft}
\begin{tabular}{lcrr}
\hline\\
Asteroid&Span&No. Obs.&L\\
\smallskip\\
\hline\\
(588)Achilles\quad&1906-1990\quad&155\quad&5\\
(617)Patroclus&1906--1989&114&4\\
(624)Hector&1906--1990&237&5\\
(659)Nestor&1908--1990&91&5\\
(884)Priamus&1917--1989&95&4\\
(911)Agamenon&1919--1984&124&5\\
(1143)Odysseus&1930--1990&164&5\\
(1172)Aneas&1930--1987&112&4\\
(1173)Anchises&1930--1989&76&4\\
(1208)Troilus&1931--1989&45&4\\
(1404)Ajax&1936--1989&60&5\\
(1437)Diomedes&1937--1989&110&5\\
\smallskip\\
\hline
\end{tabular}
\end{flushleft}
\bigskip
\end{table}

\begin{table*}
\caption{Reference orbit initial conditions. the columns show the asteroid
number, coordinate and velocity of the asteroid, referred to the equator and
equinox of 1950.0 at the epoch 2448600.5JD.}
\label{ConIni}
\begin{flushleft}
\begin{tabular}{lccc}
\hline\\
Ast&$x$&$y$&$z$\\
   &$\dot{x}$&$\dot{y}$&$\dot{z}$\\
\smallskip\\
\hline\\

588 & -3.9025983497742 & -3.2736728834505 & -2.5316108761226\\
    &  0.0043537492285 & -0.0047928164771 & -0.0022300129791\\
617 &  -1.0627928190858  &   3.9020966430745  &   3.6159794283693\\
    &  -0.0069836057153  &  -0.0015890388164  &   0.0011178570651\\
624 &  -3.0501202747117  &  -2.9280029114512  &  -2.9211148785811\\
    &   0.0059844244267  &  -0.0039228257827  &  -0.0026266504514\\
659 &  -5.4934657772575  &  -1.0458211881904  &  -0.6352066069516\\
    &   0.0020841588998  &  -0.0058962922388  &  -0.0030907056851\\
884 &   0.1672107459144  &   5.0820722624050  &   2.7350374790094\\
    &  -0.0066515083585  &   0.0006701997987  &  -0.0006529235805\\
911 &  -3.5738399893703  &  -2.6976481338099  &  -3.2968414383167\\
    &   0.0052455725404  &  -0.0038870724565  &  -0.0026647074728\\
1143&  -4.0433167628393  &  -3.4378827005772  &  -1.4783781326443\\
    &   0.0044210292769  &  -0.0052935067994  &  -0.0018697093635\\
1172&  -0.6758109101219  &   5.4416321288771  &   1.4386284702578\\
    &  -0.0065958884083  &  -0.0000789213947  &  -0.0019103282979\\
1173&  -1.5913219103482  &   5.1937286624775  &   2.2182587621419\\
    &  -0.0065720689729  &  -0.0007955518902  &  -0.0012263064683\\
1208&  -1.2576214476810  &   3.1558146702659  &   4.2762472615874\\
    &  -0.0064452466474  &  -0.0029968982314  &   0.0010308113922\\
1404&  -4.3891545802031  &  -2.5604362090287  &  -2.9392856669528\\
    &   0.0044233909857  &  -0.0042518492743  &  -0.0026720722752\\
1437&  -2.8517339598160  &  -3.0129925076262  &  -3.3137589975109\\
    &   0.0058058535762  &  -0.0042204981599  &  -0.0014436102720\\
\smallskip\\
\hline\\
\end{tabular}
\end{flushleft}
\end{table*}

\section{Data set and Ephemeris}

           As discussed in the previous work, only those asteroids with a
minimum observation time of 60 years were selected, which is required to
determine the mass correction of Saturn. An error in the Saturn mass
introduces a spurious shift in the Lagrange equilibrium point. At present,
six new Trojan asteroids can be incorporated giving rise to twelve asteroids
satisfying that condition.

           The observations were obtained from the MPC (after 1950) and from
several publications (before 1950) from 1906 to 1990. A total amount of 1383
observations were employed (see Table \ref{Param}), which were reduced to
astrometric position and the observation time to ephemeris time. This
observational material will be separately published.

           The equations of motion of each Trojan asteroid were numerically
integrated, together with those of the outer planets, using a heliocentric
coordinate system referred to the equinox and equator of 1950.0. A standard
predictor-corrector of fifth order with a step of five days was used for the
numerical integration. The positions of the outer planets were tested with
the standard ephemeris (Eckert, 1951). Besides, the variational equations
for each asteroid and Jupiter were simultaneously integrated.

\section{The new results}

          The initial conditions for each asteroid were obtained from the
Ephemeris for Minor Planets 1991, for the epoch 2448600.5JD, and were
adjusted through a differential correction in order to obtain a reference
orbit. Table \ref{ConIni} shows the initial conditions for the reference
orbit of each asteroid and Table \ref{Param} shows the remainding
observations once those with residuals larger than 3 were eliminated (Arley,
1950).

          By using this reference orbit, several adjustments were carried out
for each asteroid. The obtained results are illustrated in Table \ref{SolInd}
and \ref{SolSim}. Table \ref{SolInd} shows the separate and the joint
solutions of the inverse of the Saturn mass and the Nordtvedt parameter with
the orbital elements of each asteroid.

\begin{table*}
\caption[Mass correction and the Nordtvedt parameter: preliminary results]%
        {Mass correction and the Nordtvedt parameter: preliminary results.
The columns show the asteroid number, the value of Saturn inverse mass, the
Nordtvedt parameter (obtained from separate fits to each asteroid) and the
values obtained fitting both parameters to each asteroid.}
\label{SolInd}
\begin{flushleft}
\begin{tabular}{lcccc}
\hline\\
        &\multicolumn{2}{c}{Separate fits}&\multicolumn{2}{c}{Joint fit}\\
\smallskip\\
\hline\\
Ast     & $M_\odot/M_S$ &$ \eta$ & $M_\odot/M_S$ &$ \eta$       \\
\smallskip\\
\hline\\
588     & $3499.83\pm0.88 $ & $-0.37\pm0.18$ & $3499.43\pm0.89$ &
                                                        $-0.45\pm0.18$ \\
617     & $3489.83\pm1.26 $ & $ 0.60\pm0.15 $ & $3488.57\pm1.57$ &
                                                        $-0.22\pm0.17$  \\
624     & $3502.26\pm1.59 $ & $-0.80\pm0.14$ & $3497.64\pm1.71$ &
                                                        $-0.97\pm0.16$  \\
659     & $3499.72\pm1.01 $ & $ 0.31\pm0.16$ & $3498.97\pm1.24$ &
                                                        $-0.20\pm0.19$  \\
884     & $3499.38\pm1.10 $ & $ 0.26\pm0.24$ & $3499.35\pm1.33$ &
                                                        $-0.01\pm0.28$  \\
911     & $3510.63\pm4.13 $ & $-0.61\pm0.33$ & $3508.43\pm5.15$ &
                                                        $-0.29\pm0.40$  \\
1143    & $3495.86\pm1.17 $ & $-0.82\pm0.18$ & $3496.38\pm2.69$ &
                                                        $-0.09\pm0.42$  \\
1172    & $3496.84\pm0.63 $ & $ 1.05\pm0.16$ & $3498.19\pm0.77$ &
                                                        $ 0.57\pm0.19$  \\
1173    & $3494.77\pm1.06 $ & $ 1.45\pm0.34$ & $3493.93\pm1.66$ &
                                                        $-0.32\pm0.50$  \\
1208    & $3487.86\pm5.39 $ & $-1.39\pm1.01$ & $3474.18\pm10.3$ &
                                                        $2.97\pm1.92$   \\
1404    & $3497.29\pm2.06 $ & $-0.01\pm0.26$ & $3491.20\pm3.15$ &
                                                        $1.07\pm0.43$   \\
1437    & $3489.73\pm2.30 $ & $-0.14\pm0.29$ & $3489.69\pm2.32$ &
                                                        $0.03\pm0.28$   \\
\smallskip\\
\hline\\
\end{tabular}
\end{flushleft}
\end{table*}

          The values of the inverse of the Saturn mass are in rough
agreement with those obtained by other researches, although their dispersion
is greater than the formal errors. On the other hand, the values of the
Nordtvedt parameter $\eta$ show a bias toward negative values and is strongly
correlated with the L$_4$, L$_5$ positions.

	  This fact suggests the existence of systematic errors in the
reference system of the star catalogues. In order to analize this effect, a
simultaneous adjustent of all 12 asteroids was made. A total of 77
parameters were adjusted, including the correction to the Saturn mass,the
Nordtvedt parameter and three parameters that describe biases of the star
catalogue reference system.

          The correction for the Saturn mass is in good agreement with the
value recommended by the IAU with the corrections suggested by the JPL
results. The corrections to the star catalogue reference system are in
agreement with other determinations. The value of the Nordtvedt parameter
$\eta$ is not zero at the 95\% confidence level.

\begin{table*}[t]

\caption[Final values of the physical parameters]%
        {Final values of the physical parameters, as obtained from a
simutaneous fit of all asteroids. The columns show the supressed asteroid
number, and the adjusted values of the inverse mass of Saturn, the Nordtvedt
parameter and the three astrometric bias parameters. The last line contains
the ``jackknifed'' values of the parameters}
\label{SolSim}
\begin{flushleft}
\begin{tabular}{lccccc}
\hline\\
Ast     & $M_\odot/M_S$ &$ \eta$ & $\Delta\psi$ & $\Delta\phi$ &
                                                        $\Delta{}b$     \\
\smallskip\\
\hline\\
---     & $3497.78\pm0.40$ & $-0.17\pm0.07$ & $0.84\pm0.12$ & $-0.19\pm0.14$
                                                & $-0.08\pm0.05$        \\
588     & $3497.20\pm0.46$ & $-0.25\pm0.07$ & $0.94\pm0.12$ & $-0.39\pm0.15$
                                                & $-0.04\pm0.05$        \\
617     & $3498.42\pm0.42$ & $-0.18\pm0.09$ & $0.81\pm0.14$ & $-0.21\pm0.16$
                                                & $-0.06\pm0.05$        \\
624     & $3497.76\pm0.41$ & $-0.11\pm0.07$ & $0.67\pm0.13$ & $-0.19\pm0.16$
                                                & $-0.13\pm0.05$        \\
659     & $3497.72\pm0.44$ & $-0.21\pm0.07$ & $0.88\pm0.12$ & $-0.22\pm0.15$
                                                & $-0.05\pm0.05$        \\
884     & $3497.55\pm0.43$ & $-0.16\pm0.07$ & $0.85\pm0.12$ & $-0.18\pm0.15$
                                                & $-0.04\pm0.05$        \\
Q911     & $3497.66\pm0.39$ & $-0.15\pm0.07$ & $0.77\pm0.11$ & $-0.10\pm0.15$
                                                & $-0.12\pm0.05$        \\
1143    & $3497.70\pm0.45$ & $-0.19\pm0.07$ & $0.85\pm0.12$ & $-0.22\pm0.15$
                                                & $-0.06\pm0.05$        \\
1172    & $3497.79\pm0.44$ & $-0.19\pm0.07$ & $0.82\pm0.12$ & $-0.19\pm0.15$
                                                & $-0.10\pm0.05$        \\
1173    & $3497.84\pm0.42$ & $-0.13\pm0.07$ & $1.03\pm0.13$ & $-0.29\pm0.15$
                                                & $-0.10\pm0.05$        \\
1208    & $3497.87\pm0.41$ & $-0.15\pm0.07$ & $0.81\pm0.12$ & $-0.11\pm0.15$
                                                & $-0.06\pm0.05$        \\
1404    & $3497.71\pm0.42$ & $-0.19\pm0.07$ & $0.80\pm0.12$ & $-0.12\pm0.15$
                                                & $-0.08\pm0.05$        \\
1437    & $3497.85\pm0.42$ & $-0.18\pm0.07$ & $0.85\pm0.12$ & $-0.16\pm0.15$
                                                & $-0.09\pm0.05$        \\
Jack    & $3497.80\pm0.81$ & $-0.14\pm0.12$ & $0.75\pm0.31$ & $-0.29\pm0.32$
                                                & $-0.16\pm0.10$        \\
\smallskip\\
\hline\\
\end{tabular}
\end{flushleft}
\end{table*}

\section{Conclusions}

        On the data shown in Tables \ref{SolInd} and \ref{SolSim} are based
the main results of this paper. Let us discuss them in some detail.
        The dispersion of the inverse of the Saturn mass and the Nordtvedt
parameter in Table \ref{SolInd} suggest the existence of unmodelled
systematic errors and, in particular, the strong correlation of $\eta$ with
the preceding or receding position of the asteroid with respect to Jupiter,
shows the influence of the equinox shift in the star catalogue reference
system on our results. This latter quantity, indeed, must introduce spurious
shifts $\delta\eta$ in the Lagrange points $L_4$ and $L_5$ but with opposite
signs.  The same should be true for an error in the Saturn mass.

        We can confirm the above comments with a simple reduction of the
$\eta$ and $M_S$ values in Table \ref{SolInd}. We use the following equation
to model the mass and equinox errors:
\begin{eqnarray}
M_S & = & {M_S}_0 + \tau {M_S}_1,                 \label{Mod1Err}      \\
\eta & = & \eta_0 + (M_S-3500)\eta_1 + \tau\eta_2,\label{Mod2Err}
\end{eqnarray}
where $\tau=\pm1$ according to the preceding or receding position of the
asteroid with respect to Jupiter. We obtain
\begin{eqnarray}
{M_S}_0 & = & 3497.67\pm0.63    \\
\eta_0 & = & -0.08\pm0.20
\end{eqnarray}
and these values are in very good agreement with other determinations.  The
very simple model (Eqs. (\ref{Mod1Err}) and
 (\ref{Mod2Err})) takes into account very well the main systematic errors in
our determinations.

        On the other hand, the results in the first line of Table
\ref{SolSim} come from a similar but more complex (and more realistic) model
for systematic errors. The non zero value of $\eta$ is at variance with the
LLR results, and one should attempt a more rigorous estimate of the errors.

        A robust procedure for error estimation is the so called {\em
jackknife\/} process (Kinsella, 1986 ; Miller, 1974). It is a rigorous
generalization of the well known process of discarding part of a data set to
test the sensibility of a result computed from it. Let $\eta_0$ be the value
of $\eta$ computed from the full data set of size $N$, and let $\eta_i$ be
the values obtained deleting the set of $n_i$ obsevations corresponding to
asteroid $i$ from the data set. The Jackknife process consists in forming
the {\em pseudovalues\/}:
\begin{equation}
\eta^*_i = {{N\eta_0 - (N-n_i)\eta_i}\over{n_i}}
\end{equation}
which are treated as independent identically distributed random variables
and the value and error of $\eta$ are computed from them.  As a general
rule, jackknife error estimates are larger than least squares estimates,
since the process of forming pseudovalues enhances nongaussian contributions
to the dispersion, such as introduced by nonlinearities and local
distortions in star catalogues.

        The jackknife procedure, applied to the value of Table \ref{SolSim},
yields the results quoted in the last line of the table.  All the jackknife
errors estimates are about twice as big as the least squares error
estimates, probably due to the existence of unmodelled systematic errors in
the data set.  The jackknifed estimate of $\eta$ is now consistent with
zero.  As we have mentioned in (Orellana-Vucetich, 1988) the gravitational
energy of the Sun is overestimated in Eq. (\ref{Omega}) by a factor 4. The
final result for the Nordtvedt parameter in this paper is:
\begin{equation}
\eta = -0.56\pm0.48,                     \label{etafinal}
\end{equation}
 consistent with zero and not very different from our former
result (Orellana-Vucetich, 1988) . Our Eq. (\ref{etafinal}), however,
includes a rigorous jackknife estimate of external errors.

        Our results confirm the possibility of obtaining good estimates of
the Nordtvedt parameter from the motion of the Trojan asteroids. Both the
simple model of Eqs. (\ref{Mod1Err}) and (\ref{Mod2Err}) and our more
elaborate jackknife analysis show that most of the errors come from the
catalog reference system biases. A new reduction of the available plates
with respect to the forthcoming Hipparcos catalog should yield an estimate
of $\eta$ ten times as accurate as Eq. (\ref{etafinal}), of the order of
magnitude of LLR determinations. Since the Nordtvedt effect is a strong test
of General Relativity (Will, 1981), we think it is worth the efforts to test
for its existence.

\section*{Acknowledgements}

        We are grateful to G. Marsden for making available the MPC
collection of observations and to S. Ferraz-Mello for discussions and
encouragement. This work has been partially supported by PROFICO (CONICET).


\begin{thebibliography}{}

\bibitem[1950]{Arley} Arley, N., Buch, K.R.: 1950, {\em Probability and
   Statistics}, Wiley, New York.

\bibitem[1951]{Eckert} Eckert, W.J., Brouwer, D., Clemence, G.M.: 1951,
{\em Astron. Papers Amer. Eph.}, {\bf 12}

\bibitem[1986]{Kinsella}Kinsella, A.: 1986, {\em Am. J. Phys.\ }{\bf 54}, 454

\bibitem[1974]{Miller}Miller, R.G.: 1974, {\em Biometrica\ }{\bf 61}, 1

\bibitem[1968]{Nordtvedt} Nordtvedt, K.: 1968, {\em Phys.Rev.}{\bf169}, 1014

\bibitem[1988]{Orellana-Vucetich}Orellana, R.B. and Vucetich, H.: 1988,
 {\em Astron. Astrophys.\ }{\bf 200}, 248

\bibitem[1981]{Will} Will, C.M.: 1981, {\em Theory and Experiment in
 Gravitational Physics}, Cambridge University Press, Cambridge


\end{thebibliography}
\end{document}